\documentclass{article}

\usepackage{bm}
\usepackage{stmaryrd, mathrsfs, dsfont, amsmath, latexsym}

\usepackage{graphics,lscape,fancyhdr,array,euscript}
\usepackage{comment}
\usepackage[compat=1.1.0]{tikz-feynman}
\usepackage{tikz}
\usetikzlibrary{calc,decorations.markings}
\usepackage{jheppub}
\usepackage{hyperref}
\usepackage{footmisc}
\usepackage{tikz-feynman}
\usepackage{tikz-3dplot}
\usepackage{wasysym}

\usepackage{orcidlink}

\newcommand{\ba}{\mathbf{a}}
\newcommand{\bb}{\mathbf{b}}
\newcommand{\bc}{\mathbf{c}}
\newcommand{\be}{\mathbf{e}}
\newcommand{\bbf}{\mathbf{f}}
\newcommand{\bg}{\mathbf{g}}
\usepackage[utf8]{inputenc}
\usepackage{mathptmx}      
\usepackage{flushend}
\usepackage[numbers,sort&compress]{natbib}
\title{Differential Equations for Dressed $in\!-\!in$ Correlators}
\author{Ivo Sachs\orcidlink{0000-0003-2526-8655}} 

\emailAdd{ivo.sachs@lmu.de}

\affiliation[]{Arnold-Sommerfeld-Center for Theoretical Physics, Ludwig-Maximilians-Universit\"at M\"unchen, 
Theresienstr. 37, 80333 Munich, Germany}
\date{\today}

\abstract{
 We extend the formalism of differential equations and IBP identities, to cosmological $in\!-\!in$ correlators of conformally coupled scalar fields, that can be formulated as dressed flat space amplitudes. These differential equations turn out to be quite powerful for extracting information about the class of functions which the corresponding amplitudes integrate to. Furthermore, they can be formulated algorithmically. Here we illustrate this for the cosmological bubble- triangle and exchange diagram.}

\begin{document}

\maketitle

\section{Introduction}
Differential equations have been instrumental, not just to understand the structure, but also to help   compute Feynman integrals through isolating their dependence on external kinematics and masses. For a review and references see eg. \cite{Henn:2014qga}. 

On another front, there has recently been much focus on the calculation of cosmological, or $in-in$ correlations functions at future infinity in deSitter space-time. The canonical approach to such correlation functions uses a Schwinger-Keldysh (SK) contour. Differential equations have previously been obtained for tree-level wave function coefficients \cite{Arkani-Hamed:2023kig} as well as for some 1-loop diagrams at the integrand level \cite{He:2024olr} as well as for 2- and 3-site loop integrals \cite{Benincasa:2024ptf}. Similarly, differential equations for massless and massive correlators at tree-level \cite{Liu:2024str,Li:2026xhw} and 1-loop level \cite{Baumann:2024mvm,Chen:2026dqp} have been derived in the literature.

Some cosmological correlation functions, particularly for conformally coupled fields, can be equivalently expressed as flat space scattering amplitudes, dressed by auxiliary energy propagators \cite{Chowdhury:2025ohm,Chowdhury:2023arc}. The resulting integrals are structurally close to flat space higher loop amplitudes, dressed by auxiliary energy loop integrals. While these integrals cannot possibly feature Lorenz invariance, it is still possible to apply the machinery of flat space differential equations, albeit with an important modifications: Since, the auxiliary propagators are often 1-dimensional, we can use partial fraction throughout the integration by parts \cite{Chetyrkin:1981qh} (IBP)-reduction which considerably reduces the complexity of these higher loop integrals. On the other hand, auxiliary integrals may be over a finite range so that new boundary terms can appear during the IBP reduction.  

In this paper we will consider cosmological correlation functions at future infinity of a conformally coupled scalar in deSitter space. This is the closest to a massless scalar field in Minkowski space. We the extend the the IBP reductions to such diagrams and derive the connection matrix for some important cases. As we will see, in analogy to flat space amplitudes \cite{Henn:2013pwa,Primo:2016ebd,Forner:2026vby},  the leading singularity of the flat space (inner) amplitudes together with the "outer" IBP for the auxiliary propagators already determines the top sector of the connection matrix and thereby the nature, or transcendentality of the final integral. As mentioned above, differential equations can also be applied directly to the SK-integral  \cite{Arkani-Hamed:2023kig,Benincasa:2024ptf,He:2024olr,Liu:2024str,Li:2026xhw,Baumann:2024mvm,Chen:2026dqp}. What distinguishes our formulation is its proximity to scattering amplitudes with all the algorithmic tools available for it. 
We will illustrate this with several examples, including the cosmological triangle diagram. 

\section{Dressing rules}
Cosmological correlation functions in deSitter space-time  feature translation invariance in space but not in time. By consequence, energy is not conserved at the interaction  vertices. In the usual Schwinger-Keldysh representation of a perturbative diagram one is then left with nested time-integrals of vertex function of a sum over Feynman-, anti-Feynman- and Wightman propagators resulting in a less compact representation of perturbation theory. The dressing rules \cite{Chowdhury:2025ohm} is an attempt to recover a form that is close to flat space perturbation theory, by first performing the sum over SK-propagators and then trading the remaining time integrals for some auxiliary energy integrals through a Laplace-transform. 
While such rules can be formulated for any field \cite{GNSV}, they become particularly simple for a conformally coupled scalar field with Lagrangian 
\begin{align}
    L =  \frac{1}{2} g^{\mu\nu} \partial_{\mu} \phi \partial_{\nu} \phi - \frac{1}{6} R \phi^2 
    -V(\phi)~. 
\end{align}
In this paper we will focus on this case. The dressing rules can then be summarized as follows \cite{Chowdhury:2025ohm}:
\begin{enumerate}
    \item Write down a flat space Feynman diagram and relax energy conservation, $p_{\rm tot}=0$.
    \item Attach an auxiliary propagator to each vertex and set the energy of each auxiliary propagator to be $p_{\rm tot}$ of the vertex it is attached to. In general there will be two types of auxiliary propagators. We will denote the two types using dashed (that come in even numbers) and dotted lines (that come in any number), respectively. Concretely, 
     \begin{align}
        \phi^4:&\qquad \begin{tikzpicture}[baseline=($(b)!0.25!(w)$.base)]
            \begin{feynman}
                \vertex (a);
                \vertex [ right = 50pt of a ,   dot   ] (b) {};
                \vertex [ right = 40pt  of b](w);
                \diagram* {
                    (b)  -- [momentum= $p_{\rm tot}$, scalar] (w),
                };
            \end{feynman}
        \end{tikzpicture} =-\frac{2k_{\rm ext}}{p_{\rm tot}^2+k_{\rm ext}^2}\,,
        \label{eq:CC4AuxProp}\\
        \phi^3:&\quad \begin{tikzpicture}[baseline=(b.base)]
            \begin{feynman}
                \vertex (a);
                \vertex [ right = 50pt of a, dot      ] (b) {};
                \vertex [ right = 40pt  of b](w);       
                \diagram* {
                    (b)  -- [momentum= $p_{\rm tot}$, scalar] (w),
                };
            \end{feynman}
        \end{tikzpicture} 
        \label{eq:CC3UnsuppAuxProp}
    =-2i \int_0^\infty ds\, \frac{p_{\rm tot}}{p_{\rm tot}^2+(s+k_{\rm ext})^2}\,,\quad \begin{tikzpicture}[baseline=($(b)!0.25!(w)$.base)]
            \begin{feynman}
                \vertex (a);
                \vertex [ right = 50pt of a , dot      ] (b){};
                \vertex [right = 50pt of b        ] (c);
                \vertex [above left = 35.36pt and 35.36pt of b](oi);
                \vertex [below left = 35.36pt and 35.36pt of b](oj) ;
                \vertex [below = 7pt of b](z) ;
                \vertex [ right = 40pt of b](w);
                \diagram* {
                    (b)  -- [momentum= $p_{\rm tot}$, ghost] (w),
                };
            \end{feynman}
        \end{tikzpicture} = -\pi \nonumber
    \end{align}
  \item  Sum over all ways of attaching the two types of auxiliary propagators subject to some constraints that we specify below.
    \item Integrate over all the auxiliary energy variables.
\end{enumerate}
We may attach the loose end of all auxiliary propagators to a single point, which we denote below by $\Circle$. This makes the resulting diagram look identical to a higher loop flat space-diagram with local energy conservation. Note that the propagators in \eqref{eq:CC4AuxProp} are sensitive to the choice of an UV-regulator \cite{Chowdhury:2025ohm}. We will comment on that when appropriate. 

The diagrams considered in this paper are the four-point 1-loop bubble, the six-point 1-loop triangle, as well as the tree-level exchange diagram in $\lambda\phi^3$ theory, 
\begin{align}
\begin{tikzpicture}[baseline]
\draw (0,0) ellipse (1 and 0.5);
\draw[dashed] (1,0) -- (-1, 0);
\draw (1,0) -- (1.4, 0.4);
\draw (1,0) -- (1.4, -0.4);
\draw (-1,0) -- (-1.4, 0.4);
\draw (-1,0) -- (-1.4, -0.4);
\node at (-0,0) {$\Circle$};
\node at (1,0) {$\bullet$};
\node at (-1,0) {$\bullet$};
\end{tikzpicture}\qquad\quad\begin{tikzpicture}[baseline]
\begin{scope}
    \draw [dashed, postaction={decorate}](0,0.25) -- (0,-0.5);
    \draw[dashed,postaction={decorate}] (1,1) -- (0,0.25);
\draw[dashed,postaction={decorate}] (0,0.25) -- (-1,1);
\draw[postaction={decorate}] (0.5,-1) -- (0,-0.5);
\draw[postaction={decorate}] (-0.5,-1) -- (0,-0.5);
\draw[postaction={decorate}] (-1.5,1.5) -- (-1,1);
\draw[postaction={decorate}] (-1.5,0.5) -- (-1,1);
\draw[postaction={decorate}] (1,1) -- (1.5,1.5);
\draw[postaction={decorate}] (1,1) -- (1.5,0.5);
\draw[postaction={decorate}] (-1,1) -- (1,1);
\draw[postaction={decorate}] (1,1) -- (0,-0.5);
\draw[postaction={decorate}] (0,-0.5) -- (-1,1);
\node at (-0,0.3) {$\Circle$};
\node at (1,1) {$\bullet$};
\node at (-1,1) {$\bullet$};
\node at (0,-0.5) {$\bullet$};
\end{scope}
\end{tikzpicture}\qquad
\quad\begin{tikzpicture}[baseline]
\draw (-0.75, 0.5) -- (-0.5, 0);
\draw (-0.75, -0.5) -- (-0.5, 0);
\draw (0.75, 0.5) -- (0.5, 0);
\draw (0.75, -0.5) -- (0.5, 0);
\draw (-0.5,0) -- (0.5,0);
\draw[dashed] (0.5, 0) -- (0, -0.75);
\draw[dashed] (0, -0.75) -- (-0.5, 0);
\node at (-0,-0.75) {$\Circle$};
\node at (-0.5,0) {$\bullet$};
\node at (0.5,0) {$\bullet$};
\end{tikzpicture} \,.\nonumber
\end{align}
As dressed cosmological correlators, each of these diagrams has its peculiarity. The bubble illustrates how the formalism works in the presence of UV-divergencies. The triangle, while simple in flat space, turns out to be a highly non-trivial Feynman integral to calculate \cite{Pimentel:2026kqc} but, surprisingly results in the same class of functions as in flat space. As we will see, the differential equations make this manifest without actually having to calculate the integral. 
Finally, the tree-level exchange features a new class of auxiliary propagators that will also arise in non-conformally coupled theories \cite{Chowdhury:2025ohm}. As such, the examples treated here, while relatively simple cover a broad range of different settings for the dressing rules. We will start with the bubble in the next section.

\section{Cosmological Bubble}
Using the dressing rules in the last section we can cast the Schwinger-Keldysh, $in-in$ correlator in the form of a dressed flat space diagram. In particular,  
the  s-channel contribution to cosmological 4-point bubble is given by the diagram
\begin{align}
\begin{tikzpicture}[baseline]
\draw (0,0) ellipse (1.5 and 0.7);
\draw[dashed] (1.5,0) -- (-1.5, 0);
\draw (1.5,0) -- (2.5, 1);
\draw (1.5,0) -- (2.5, -1);
\draw (-1.5,0) -- (-2.5, 1);
\draw (-1.5,0) -- (-2.5, -1);
\node at (-0,0) {$\Circle$};
\node at (-1.5,0) {$\bullet$};
\node at (1.5,0) {$\bullet$};
\node at (0.7,0.25) {$p^2+x_2^2$};
\node at (-0.7,0.25) {$p^2+x_1^2$};
\node at (0.,1) {$L^2$};
\node at (0.,-1) {$(L+Q)^2$};
\node at (-0.7,0.25) {$p^2+x_1^2$};
\node at (-2.8,1.3) {$\vec{k}_1$};
\node at (-2.8,-1.3) {$\vec{k}_2$};
\node at (2.8,1.3) {$\vec{k}_3$};
\node at (2.8,-1.3) {$\vec{k}_4$};
\node at (2.8,1.3) {$\vec{k}_3$};
\node at (2.8,-1.3) {$\vec{k}_4$};
\end{tikzpicture}\,
\end{align}
with $Q=(p,\vec y=\vec{k}_{1}+\vec{k}_{2})$. On the physical sheet, $x_1=|\vec{k}_{1}|+|\vec{k}_{2}|$ and $x_2=|\vec{k}_{3}|+|\vec{k}_{4}|$ are positive. Below we will consider $x_1$ and $x_2$ as independent variables which we will assume to be positive. However, complexification of $x_1$ \cite{Chowdhury:2026upp} can easily be incorporated at the price of having some square roots in the formulae. The Feynman integral, in dimensional regularization 
 \begin{align}\label{eq:bubnc}
 C_4(x_1,x_2,\vec{y})=  
 &\int\limits_{-\infty}^\infty \frac{dp}{2\pi} \int \frac{dL^{4-2\epsilon}}{(2\pi)^4}\frac{x_1x_2}{(x_1^2+p ^2)(x_2^2+p ^2)}\frac{ \mu^{2\epsilon}}{L^2(L+Q)^2}\,
\end{align}
can be integrated in closed form \cite{Chowdhury:2023arc} to give 
\begin{equation}
C_4=-\frac{1/\epsilon}{x_1+x_2}+\frac{1}{(2\pi)^{2}}\frac{1}{16}\left[\frac{1}{x_1+x_2}\ln\left(\frac{(x_1+y)(x_2+y)}{4\mu^{2}}\right)+\frac{1}{x_1-x_2}\ln\left(\frac{x_2+y}{x_1+y}\right)\right]+O(\epsilon)\,,
\label{1loop4ptcutoff}
\end{equation}
where $y=|\vec y|$. Note that \eqref{1loop4ptcutoff} is not deSitter invariant due to the appearance of a scale $\mu$, originating from dimensional regularization and which is not compatible with the scale-invariance of boundary dS-correlators. This is so because the standard dimensional regularization employed here breaks deSitter invariance. This can be remedied using a suitable analytic regulator \cite{Chowdhury:2023arc}. This results in the replacement $\mu\to x_1+x_2$ which is just a boundary condition for the differential equation that we will derive below. So, we may stick to dimensional regularization for deriving the differential equation.

\subsection{Differential equations}
We will describe two applications of differential equations. First, in this subsection, as an illustration we will simply apply differential equations to fix the relative coefficients between the three structures in \eqref{1loop4ptcutoff}. In the second part we will instead derive a basis of master integrals for that correlator. 

The flat space bubble sub-topology in \eqref{eq:bubnc} satisfies the scaling equation (e.g. section 6 of \cite{Argeri:2007up}) 
\begin{align}\label{eq:scale}
    (p\partial_p +y\partial_y)\int  \frac{ d^{4-2\epsilon}L}{L^2((L_0+p)^2+(\vec{L}+
y \vec{e}_{3})^2)}=-2\epsilon\int  \frac{ d^{4-2\epsilon}L }{L^2((L_0+p)^2+(\vec{L}+
y\vec{e}_{3})^2)}\,,
\end{align}
where $y=|\vec{y}|$.  Combining this with the IBP-identity
\begin{align}
    0=&\int\limits_{-\infty}^\infty \partial_p p\left(\frac{x_1x_2}{(x_1^{2}+p^2)(x_2^{2}+p^2)}
 \int d^{4-2\epsilon}L \frac{ 1}{L^2((L_0+p)^2+(\vec{L}+
y\vec{e}_{3})^2)}\right)dp\nonumber\\
=&I-\int\limits_{-\infty}^\infty \frac{x_1x_2\left(\frac{2p^2}{x_1^2+p^2}+\frac{2p^2}{x_2^2+p^2}\right)}{(x_1^{2}+p^2)(x_2^{2}+p^2)}
 \int d^{4-2\epsilon}L \frac{ 1}{L^2((L_0+p)^2+(\vec{L}+
y\vec{e}_{3})^2)}dp\nonumber\\&+\int\limits_{-\infty}^\infty \frac{x_1x_2}{(x_1^{2}+p^2)(x_2^{2}+p^2)}
p\partial_p  \int d^{4-2\epsilon}L \frac{ 1}{L^2((L_0+p)^2+(\vec{L}+
y\vec{e}_{3})^2)}dp
\,,
\end{align}
we derive, in turn 
\begin{align}
(x_1\partial_{x_1}+x_2\partial_{x_2}+1)C_4=-(2\epsilon+y\partial_y)C_4\,.
\end{align}
Note that at order $1/\epsilon$ the l.h.s. vanishes. Comparing this to the structure in \eqref{1loop4ptcutoff}, including the counter term
\begin{equation}
C_4=\frac{a/\epsilon}{x_1+x_2}+\frac{1}{(2\pi)^{2}}\frac{1}{16}\left[\frac{b}{x_1+x_2}\ln\left(\frac{(x_1+y)(x_2+y)}{4\mu^{2}}\right)+\frac{c}{x_1-x_2}\ln\left(\frac{x_2+y}{x_1+y}\right)\right]\,,
\label{1loop4ptcutoffabd}
\end{equation}
this fixes $a=-b\frac{1}{(2\pi)^{2}}\frac{1}{16}$ at $O(\epsilon^0)$ but does not fix $c$.\footnote{Had we chosen the dS- a.k.a. scale-invariant regularization instead, then this equation would be trivially satisfied.  } 

Instead of eqn, \eqref{eq:scale} we may the use Lorentz-invariance of the flat space bubble
\begin{align}\label{eq:rot}
    (y\partial_p -p\partial_y)\int \frac{  d^{4-2\epsilon}L}{L^2((L_0+p)^2+(\vec{L}+
y \vec{e}_{3})^2)}=0\,.
\end{align}
Combining this with \eqref{eq:scale} we obtain
\begin{align}\label{eq:scale_rot}
    \left((y^2+p^2)\partial_y+2\epsilon y\right)\int  \frac{  d^{4-2\epsilon}L}{L^2((L_0+p)^2+(\vec{L}+
y \vec{e}_{3})^2)}=0\,,
\end{align}
and, inserting into the $p$-integral gives 
\begin{align}\label{eq:scale_rot+p}
    \left((y^2-x_1^2)\partial_y+2\epsilon y\right)C_4 =-\partial_y\int\limits_{-\infty}^\infty \frac{x_1x_2}{(x_2^{2}+p^2)}
 \int  \frac{ d^{4}L}{L^2((L_0+p)^2+(\vec{L}+
y\vec{e}_{3})^2)}dp
\,,\end{align}
where we set $\epsilon=0$ on the r.h.s. since the residue of the $\epsilon$-pole is independent of $y$, in agreement with our discussion on boundary conditions at the beginning of this section. A similar equation is obtained by the substitution $x_2\leftrightarrow x_1$. The difference between the two is, however, trivially satisfied so that only one new equation is obtained. Again, at order $1/\epsilon$, the left hand side is trivially satisfied. However, eqn. \eqref{eq:scale_rot+p} fixes the relation between $b$ and $c$. This shows, in particular, that the use of differential equations together with deSitter Ward identity fixes this four-point function completely up to the usual ambigouity in the subtraction scheme. 
\subsection{Master Integrals}
In this section we derive a basis of master integrals following the Laporta/IBP/master integral route \cite{Laporta:2000dsw} for the cosmological bubble. We start with decomposing the outer (auxiliary) sector as a partial fraction,
\begin{align}\label{eq:c4d}
C_{4}(x_1,x_2,\vec{y})=&\int\limits_{-\infty}^\infty\left(\frac{x_1x_2}{(x_1^{2}+p^2)(x_2^{2}+p^2)}
 \int \frac{  d^{4-2\epsilon}L}{L^2(L+Q)^2}\right)dp\nonumber\\
=&\frac{x_1x_2}{x_2^{2}-x_1^{2}}\int\limits_{-\infty}^\infty\left(\frac{1}{x_1^{2}+p^2}-\frac{1}{x_2^{2}+p^2}\right)
 \int \frac{ d^{4-2\epsilon}L }{L^2(L+Q)^2}dp\nonumber\\
=&:\frac{x_1x_2}{x_2^{2}-x_1^{2}} (I^\epsilon_{x_1,\vec{y}}-I^\epsilon_{x_2,\vec{y}})\,, 
\end{align}
The master integrals in the last line have the graphical representation
\begin{align}
I^\epsilon_{x,\vec{y}}=   &\quad\begin{tikzpicture}[baseline]
\draw (0,0) ellipse (1.3 and 0.65);
\draw[dashed] (1.3,0) -- (-1.3, 0);
\node at (0,0.25) {$p^2+x^2$};
\node at (1.3,0) {$\bullet$};
\node at (-1.3,0) {$\bullet$};
\end{tikzpicture}\,.
\end{align}
To obtain differential equations for $I^\epsilon_{x,\vec{y}}$ we set
\begin{align}
    T_{11}:=\int \frac{ d^{4-2\epsilon}L }{L^2(L+Q)^2}\,.
\end{align}
Then combining, Lorentz invariance  
\begin{align}
    \left(p\partial_{\vec{y}}- \vec{y}\partial_p\right)\;T_{11}=0\,
\end{align}
and scale symmetry 
\begin{align}
\left(\vec{y}\partial_{\vec{y}}+p\partial_p+2\epsilon\right)\;T_{11}=0\,,
\end{align}
we derive
\begin{align}\label{eq:idif}
\left((p^2+y^2)\partial_{y^2}+\epsilon\right)\,T_{11}=0\,.
\end{align}
Inserting this into $I^\epsilon_{x,\vec{y}}$ we get algebraically, without  using  IBP, 
\begin{align}\label{eq:paxy}
  \left((y^2-x^2)\partial^\epsilon_{y^2}+\epsilon\right)I^\epsilon_{x,\vec{y}}=-\partial_{y^2} I^\epsilon_{\vec y}\,.  
\end{align}
Here,
\begin{align}
  I_{\vec y}^\epsilon=  \int \limits_{-\infty}^\infty dp\int  \frac{ d^{4-2\epsilon}L}{L^2(L+Q)^2}\quad=  \quad &\begin{tikzpicture}[baseline]
\draw (0,0) ellipse (1 and 0.5);
\node at (-1,0) {$\bullet$};
\node at (1,0) {$\bullet$};
\end{tikzpicture}\,,
\end{align}
satisfies 
\begin{align}\label{eq:pay}
    \left(2y^2\partial_{y^2}+2\epsilon-1\right) I_{\vec y}^\epsilon=0\,.
\end{align}
So we have reached the end of the reduction where no new master integrals are produced by differentiation. A basis of master integrals is thus given by  
\begin{align}
  \bm m=  (I^\epsilon_{\vec y},I^\epsilon_{x_1,\vec{y}}, I^\epsilon_{x_2,\vec{y}})^T\,.
\end{align}
We can then read off the corresponding connection matrix from \eqref{eq:paxy} - \eqref{eq:pay} as 
\begin{align}
 \bm A=   \begin{pmatrix}
        \frac{1-2\epsilon}{2y^2}&0&0\cr\frac{2\epsilon-1}{2y^2(y^2-x_1^2)} &\frac{\epsilon}{x_1^2-y^2}&0\cr 
        \frac{2\epsilon-1}{2y^2(y^2-x_2^2)} &0&\frac{\epsilon}{x_2^2-y^2}
    \end{pmatrix}\,.
\end{align}
The integration of the equation $\partial_{y^2}\bm m=\bm A\,\bm m$ is straight forward. From the first row  we conclude
\begin{align}
    I^\epsilon_{\vec y}=c\,(y^2)^{\frac{1-2\epsilon}{2}}\,,
\end{align}
where $c$ is independent of $y$ and fixed by the boundary condition. The diagonal entry in the second line then implies 
\begin{align}
    I^\epsilon_{x_1,\vec{y}}=\frac{f(y)}{(x_1^2-y^2)^\epsilon}\,
\end{align}
with $f$ determined by $A_{21}$. Concretely, 
\begin{align}
    f(y)=\frac{c}{x_1}\text{ArcTanh}\frac{y}{x_1}+f(0)\,,
\end{align}
where $f(0)$ is again fixed by the boundary condition. Comparison with \eqref{eq:c4d} then gives $c=-2\pi^3$ and $f(0)=\pi^3\frac{\mu^{-\epsilon}}{\epsilon x_1}$, The latter depends on the subtraction scheme. So, the integral is uniquely determined up to a global, multiplicative constant and a choice of scheme. Had we chosen a dS-invariant regulator instead, as in \cite{Chowdhury:2023arc}, this would affect only the boundary condition through $\mu\to x_1+x_2$. 

\section{Cosmological Triangle}
In this section we consider the conformally coupled 1-loop six-point function in deSitter space, or cosmological triangle given by a dressef Feynman integral of the form \cite{Chowdhury:2025ohm}
\begin{align}\label{tria4D01}
& \int\limits_{-\infty}^\infty  \frac{d p_2 dp_3 }{((p_3-p_2)^2 + x_0^2)(p_2^2 + x_2^2)(p_3^2 + x_3^2)}\;\;\int  \frac{d^{4}L}{L^2 (L + Q_2)^2(L + Q_3)^2}\,. 
\end{align}
In order to derive a basis of master integrals for this diagram we will need to consider an extended family of integrals
\begin{align}\label{tria4D1}
&I^{m_1m_2m_3r_1r_2r_3;j}_{n_1n_2n_3}= \int\limits_{-\infty}^\infty [d p_2]_\delta [dp_3]_\delta \stackrel{i_{m_1m_2m_3r_1r_2r_3}}{\overbrace{\frac{1}{D_1^{m_1}D_2^{m_2}D_3^{m_3}D_4^{r_1}D_5^{r_2}D_6^{r_3}\Lambda^j }}}\;\;\stackrel{T_{n_1n_2n_3}}{\overbrace{\int  \frac{d^{4-2\epsilon}L}{D_7^{n_1} D_8^{n_2} D_9^{n_3}}}}\,,
\end{align}
where
\begin{align}
    &D_1=(p_3-p_2)^2 + x_0^2\,,\quad D_2=p_2^2 + x_2^2\,,\quad D_3=p_3^2 + x_3^2\,,\quad\nonumber\\
    &D_4=p_2p_3 + \frac{\bc}{2}\,,\quad D_5=p_2^2 + \ba\,,\quad D_6=p_3^2 + \bb\,,
\end{align}
are auxiliary propagators and 
\begin{align}\label{eq:triL}
  \Lambda=D_4^2-D_5D_6\,
\end{align}
is the Källén function originating in the flat space massive triangle amplitude $T_{111}$ with 
\begin{align}
   D_7=L^2\,,\quad D_8=(L + Q_2)^2\,,\quad D_9=(L + Q_3)^2
\end{align}
for flat space external momenta 
\begin{align}
 Q_2 = (p_2, \vec y_2), \qquad Q_3= (p_3,  \vec y_3)\,.
\end{align}
As in the previous section we take the energy of the deSitter external energies $x_i$ off-shell, i.e. independent of the $\{|\vec{k}_i|\}$. The actual cosmological triangle integral we want to compute is given by $I^{111000;0}_{111}$. The extra master integrals contained in $I^{m_1m_2m_3r_1r_2r_3;j}_{n_1n_2n_3}$ are needed to close the system of differential equations with the identifications 
\begin{align}
  \ba=\vec{y}_2^2\,,\quad
  \bb=\vec{y}_3^2\,,\quad\bc=2\vec{y}_2\cdot\vec{y}_3\,.
\end{align}

The Feynman integrals for the cosmological triangle is finite, but we nevertheless introduce a dimensional regulator $\epsilon$ for the flat space integral as well as an analytic regulators $\delta$ for the energy integrals over $p_i$ to avoid boundary terms from the IBP reduction. They can both be set to zero in the final differential equation that we will derive. Indeed, the coefficients of the three bubble subtopologies in the
$\partial_aT_{111}$ equation of 
the triangle Feynman integral 
\begin{align}
&T_{111}=\int  \frac{d^{4-2\epsilon}L}{D_7 D_8 D_9}\
\end{align}
add to zero so that their common ultraviolet pole
and scheme-dependent constant therefore cancel.  A finite basis is
\begin{equation}\label{eq:f-basis}
 \bm f=(1,L_y,L_x,T_{111})^{\mathsf T},\quad
 L_y=T_{101}-T_{011}=\log\frac{D_{10}}{D_6},\quad
 L_x=T_{110}-T_{011}=\log\frac{D_{10}}{D_5}\,,
\end{equation}
where we introduced a redundant propagator $D_{10}:=(Q_2-Q_3)^2$ for notational convenience. At fixed $p_2,p_3,\bb,\bc$, the local connection for the flat space triangle family is then 
\begin{equation}
 \partial_a\bm f=A_a^{\rm flat}\bm f,
 \qquad
 A_a^{\rm flat}=
 \begin{pmatrix}
 0&0&0&0\\
 \frac{1}{D_{10}}&0&0&0\\
 \frac{1}{D_{10}}-\frac{1}{D_{5}}&0&0&0\\
 0&\dfrac{C_y}{2\Lambda}&\dfrac{C_x}{2\Lambda}
   &\dfrac{D_6}{2\Lambda}
 \end{pmatrix},
 \label{eq:flat-connection}
\end{equation}
where
\begin{equation}
 C_y=\frac{D_4}{D_5}+\frac{D_6-D_4}{D_{10}},\qquad
 C_x=\frac{D_4-D_6}{D_{10}}.
\end{equation}
Similarly, the auxiliary $p$-integrations for the basis of master integral that we will choose will be finite for $\delta\to 0$. This finiteness property is important since, neither the $\epsilon$, nor the $\delta$ regularization preserve deSitter-invariance, so that if finite $\frac{1}{\epsilon}\cdot O(\epsilon)$ terms were to contribute they would produce non-dS invariant contributions to the final result (see, e.g. \cite{Chowdhury:2023arc,Nowinski:2025cvw}). 

From \eqref{eq:flat-connection} we can infer that one source for higher transcendentality in the cosmological  triangle comes from the top sector in the last line of the connection matrix. To explore this we will first consider the toy model
\begin{align}\label{eq:tritoy}
&\bar I^{m_1m_2m_3,r_1r_2r_3;j}= \int\limits_{-\infty}^\infty \frac{[dp_2]_\delta [dp_3]_\delta }{D_1^{m_1}D_2^{m_2}D_3^{m_3}D_4^{r_1}D_5^{r_2}D_6^{r_3}\Lambda^{j+\frac{1}{2}}}\,,
\end{align}
with $m_1,m_2,m_3,r_1,r_2,r_3,j\in \mathbf{Z}$. Eventually we are interested in $r_i=j=0$. 
One can then derive the following differential equation algebraically, with $\bar I^{n_1n_2n_3}:=\bar I^{n_1n_2n_3,000}$:
\begin{align}\label{eq:diftoy}
   \tilde\Lambda_0\partial_{\ba} \bar I^{111}=& \frac{1}{2}(\bb-x_3^2
   )\bar I^{111}+\frac{1}{2}\bar I^{110}+(\ba-\frac{\bc}{2})\partial_{\ba}\bar I^{110}+ (\bb-\frac{\bc}{2})\partial_{\ba} \bar I^{101}+\frac{\bc}{2}\partial_{\ba} \bar I^{011}\,,
   \end{align}
with similar equations for $\partial_{\bb}$ and $\partial_{\bc}$. 
Here we defined
\begin{align}
     \tilde\Lambda_0
     &=\lambda_0+\bb x_2^2+\ba x_3^2+ \frac{\bc}{2}(x_0^2-x_2^2-x_3^2)\,,
\end{align}
where $\lambda_0=\frac{\bc^2}{4}-\ba\bb$. 
This is the first step of the reduction, schematically represented as
\begin{align}
    \partial_{\ba}\begin{tikzpicture}[baseline]
\draw[dashed] (1,1) -- (0,0.5);
\draw[dashed] (0,-1) -- (0,0.5);
\draw[dashed] (0,0.5) -- (-1,1);
\draw (-1,1) -- (1,1);
\node at (-1,1) {$\bullet$};
\node at (0,0.5) {\Circle};
\node at (1,1) {$\bullet$};
\node at (0,-1) {$\bullet$};
\draw (1,1) -- (0,-1);
\draw (-1,1) -- (0,-1);
\end{tikzpicture}\quad=\quad\begin{tikzpicture}[baseline]
\draw[dashed] (1,1) -- (0,0.5);
\draw[dashed] (0,-1) -- (0,0.5);
\draw[dashed] (0,0.5) -- (-1,1);
\draw (-1,1) -- (1,1);
\node at (-1,1) {$\bullet$};
\node at (0,0.5) {\Circle};
\node at (1,1) {$\bullet$};
\node at (0,-1) {$\bullet$};
\draw (1,1) -- (0,-1);
\draw (-1,1) -- (0,-1);
\end{tikzpicture}
\quad+\quad\begin{tikzpicture}[baseline]
\draw (-1,1) -- (1,1);
\node at (-1,1) {$\bullet$};
\node at (1,1) {$\bullet$};
\node at (0,-1) {$\bullet$};
\draw (1,1) -- (0,-1);
\draw (-1,1) -- (0,-1);
\draw[dashed] (-1,1) arc(70:35:2.0cm and 5cm );
\draw[dashed] (-1,1) arc(210:330:1.1cm and 0.5cm);
\end{tikzpicture}
\quad+\quad\begin{tikzpicture}[baseline]
\draw (0,0) ellipse (1 and 0.5);
\draw[dashed] (0,1) arc(90:0:1.0cm and 1.0cm);
\draw[dashed] (0,1) arc(180:270:1.0cm and 1.0cm);
\draw[dashed] (0,1,0) -- (-1, 0);
\node at (0,1) {$\Circle$};
\node at (-1,0) {$\bullet$};
\node at (1,0) {$\bullet$};
\end{tikzpicture}\,,\label{eq:scheme}
\end{align}
where the last diagram represents sub-topologies of the flat space triangle that we drop in our toy model. They will be included later in section \ref{sec:ctr}. Note that no IBP's were needed in this step. 

The first term on the r.h.s. of \eqref{eq:diftoy} then implies that $\bar I^{111}$ has a leading singularity of the form
\begin{align}
    \bar I^{111}=\frac{1}{\sqrt{\tilde\Lambda_0}}f(\ba,\bb,\bc,x_0,x_2,x_3)\,.
\end{align}
Moreover, the fact that the only un-differentiated term on the r.h.s. of \eqref{eq:diftoy} is multiplied by the same function implies that $\bar I^{111}$ has the same transcendentality as $\bar I^{110}$, since the differential equation for $f$ does not have a simple pole. Thus we were able to reduce the maximal transendentality of $\bar I^{111}$ with little effort. However, we should note that this insight alone does not yet exclude that appearance of an elliptic sector, or non-triangularizeable block in the reduction of $\bar I^{110}$ for instance. To continue we need differential equations for $\bar I^{011}$ etc. appearing on the r.h.s. In the next subsection we illustrate this reduction for $\bar I^{110}$. 
\subsection{Reduction of $\bar I^{110}$:}
We now treat the reduction of the second diagram on the r.h.s. of \eqref{eq:scheme}. 
We already mentioned above that, in order to avoid issues with boundary terms and convergence with auxiliary subtopologies we use an analytical regulator for the  measure $[dp]_\delta=d^{1-2\delta}p_2\,d^{1-2\delta}p_3$.  Then 
\begin{align}\label{eq:l0paIb}
\partial_{\ba}\bar I^{{n_1}{n_2}0,{r_1}0{r_3};j}= &
(j+\frac{1}{2})\int\frac{[dp]_\delta \;D_6}{D_1^{n_1} D_2^{n_2} D_4^{r_1} D_6^{r_3} \Lambda^{j+\frac{3}{2}} }=(j+\frac{1}{2})\bar I^{{n_1}{n_2}0,{r_1}0{r_3}-1;j+1}\,.
\end{align}
We can now use the scaling IBP's for $p_i$ combined with partial fractions to bring the exponent $j$ back to zero. Then  
\begin{align}\label{eq:l0paIbj10}
-\lambda_0\partial_{\ba}\bar I^{{n_1}{n_2}0,{r_1}0{r_3};0}= &-\frac{\lambda_0}{2}\bar I^{{n_1}{n_2}0,{r_1}0{r_3}\!-1;1}\\
=&\left(\!1\!-\!2\epsilon\! -\!{n_1}\!-\!{r_1}\!-\!{n_2}\!-\!{r_3}\!+\!\frac{1}{2}\!\right)\bar I^{{n_1}{n_2}0,{r_1}0{r_3}\!-1;0}+{n_1}x_0^2 \bar I^{{n_1}\!+1{n_2}0,{r_1}0{r_3}\!-1;0}\nonumber\\+&\frac{{r_1}\bc}{2}\bar I^{{n_1}{n_2}0,{r_1}\!+10{r_3}-1;0}+{n_2}x_2^2\bar I^{{n_1}{n_2}\!+10,{r_1}0{r_3}\!-1;0}+({r_3}\!-\!1)\bb\bar I^{{n_1}{n_2}0,{r_1}0{r_3};0}\nonumber\,.
\end{align}
We now claim that a basis of master integrals is given by 
\begin{align}\label{eq:basis_11}
   \mathbf{m}= \begin{pmatrix}
I^{000,0}&I^{010,0}&I^{100,0}&I^{110,0}&I^{110,-1}
    \end{pmatrix}^T\,,
\end{align}
where $\bar I^{usv,t}:=\bar I^{usv,t00;0}$. To show that this we first notice that we can bring the value of $t$ back to $0$ on the r.h.s. of \eqref{eq:l0paIbj10} by use of the  $tangential$ IBP (i.e. preserving $\Lambda$) generated by the vector field 
\begin{align}
    T=(\bc p_2-2\ba p_3)\partial_{p_2}+(2\bb p_2-\bc p_3)\partial_{p_3}\,.
\end{align}
This results in 
\begin{align}
   2\lambda_0 \partial_{\ba} \bar I^{110,0}=& (\bc -3 \bb +  3 x_0^2-3x_2^2) \bar I^{110,0}-\bar I^{010,0} + \bar I^{100,0} -  2 \bar I^{110, -1}  \nonumber\\
   &+
   2 x_2^4 \bar I^{120,0} + 
   2x_2^2 ( \bar I^{020,0}  + 2 \bar I^{120, -1} + 
      (\bb - \bc - x_0^2) \bar I^{120,0} )\nonumber\\
      &+ 
   2 x_0^2 ( 
      2 \bar I^{210,-1} -\bar I^{200,0} - ( \bc -\bb + x_0^2-x_2^2) \bar I^{210,0})\,
\end{align}
and similar equations for $\partial_{\ba} \bar I^{100,0}$ etc. To continue, one-denominator Hermite identities first remove the elementary raised powers. To $0th$ order in $\delta$,
\begin{align}
     \bar I^{200,0}=& \frac{1}{2x_0^2}\bar I^{100,0}-\frac{1}{2x_0^2(\ba+\bb-\bc-x_0^2)} \bar I^{000,0}\nonumber\\
   \bar I^{020,0}=& \frac{1}{2x_2^2(x_2^2-\ba)} \bar I^{000,0}+\frac{1}{2x_2^2}\bar I^{010,0}\nonumber\\
     \bar I^{010,-1}=& \frac{\bc}{2\ba}\left(\bar I^{000,0}+(\ba-x_2^2) \bar I^{010,0}\right)\nonumber\\
      \bar I^{020,-1}=& (\frac{\bc}{2\ba}+\frac{(\ba-x_2^2)}{2x_2^2} )\bar I^{010,0}-\frac{1}{2x_2^2}\bar I^{000,0}\nonumber\\
      \bar I^{100,-1}=& -q_A\bar I^{000,0}+q_B\bar I^{100,0}\nonumber\\
      \bar I^{200,-1}=& (-q_A+\frac{q_B}{2x_0^2} )\bar I^{100,0}-\frac{q_B}{2x_0^2(\ba+\bb-\bc-x_0^2)}\bar I^{000,0}\,,
\end{align}
where
\begin{align}
    q_A&=\frac{12 \ba \bb - 4\ba \bc - 4 \bb \bc + \bc^2 }{8 (\ba + \bb - \bc)^2 }\nonumber\\
    q_B&=-\frac{(\ba + \bb - 
      \bc) (4 \ba (\bb - \bc) + (3 \bc-4 \bb) \bc) + (\bc (4 \bb - \bc) -
    4 \ba (3 \bb - \bc)) x_0^2}{8 (\ba + \bb - \bc)^2}\,.
\end{align}
A four-equation system then reduces the $v=0$ mixed integrals. At $v=-1$, three independent equations reduce the remaining numerator integrals and the fourth equation is retained as a consistency check. More details are contained in the Mathematica package [\href{https://github.com/IvoSachs/in-in_correlators-}{CosmologicalBaikov.nb}]. 

In sum we arrived at the closed set of differential equations for $I^{110}$ of the form (at $\epsilon=\delta=0$)
\begin{align}
    \partial_a \mathbf{m}=\mathbf{A}\mathbf{m}\,,
\end{align}
where $A_{ij}$ are rational functions. 
We now want to argue that it is consistent to set $\delta\to 0$. Indeed, while all elements in \eqref{eq:basis_11} except $I^{111,0}$ require regularization, none of these integrals features a pole at $\delta\to 0$. Setting $\delta\to 0$ considerably simplifies the reduction. In particular, 
\begin{align}
    \mathbf{A}=\begin{pmatrix}
        0&0&0&0&0\cr \frac{-1}{2\ba(\ba-x_2^2)}&\frac{-1}{2\ba}&0&0&0\cr
        \frac{-1}{2(\ba+\bb-\bc)((\ba+\bb-\bc)-x_0^2)}&0&
\frac{-1}{2(\ba+\bb-\bc)}&0&0\cr
 A_{41}&A_{42}&A_{43}&A_{44}&A_{45}\cr
     A_{51}&A_{52}&A_{53}&A_{54}&A_{55}
    \end{pmatrix}\,.
\end{align}
Let us pause to have a look at the first integrals of the lower rows. We find 
\begin{align}
    \bar I^{000;0}(\ba)&\sim  \bar I^{000;0}(\ba_0)\,,\quad  \bar I^{010;0}(\ba)\sim  \frac{\text{artanh}(\frac{\sqrt{\ba}}{x_2})}{
x_2\sqrt{\ba}} +\frac{c}{\sqrt{{\ba}}}\,,\quad \bar I^{100;0}(\ba)\sim \frac{\text{artanh}(\frac{\sqrt{\ba+\bb-\bc}}{x_0})}{\ba+\bb-\bc}\,,
\end{align}
which, in particular, reflects the "1-loop" nature of the energy integral 
\begin{align}
    \bar I^{010;0}=\int \frac{[d p_2]_\delta[d p_3]_\delta}{(p_2^2+x_2^2)\sqrt{p_2p_3 \bc-p_2^2\bb-p_3^2\ba+\lambda_0}}=\begin{tikzpicture}[baseline]
\draw (-1,1) -- (1,1);
\node at (-1,1) {$\bullet$};
\node at (1,1) {$\bullet$};
\node at (0,-1) {$\bullet$};
\draw (1,1) -- (0,-1);
\draw (-1,1) -- (0,-1);
\draw[dashed] (-1,1) arc(210:330:1.1cm and 0.5cm);
\end{tikzpicture}\,,
\end{align}
where is understood that the flat space triangle is reduced to its leading singularity $\frac{1}{\sqrt{\Lambda}}$. 
In order to integrate the remaining masters we then aim to diagonalize the top sector block in the next subsection.

\subsection{Diagonalization}\label{sec:diag}
 On the maximal cut $(D_1=D_2=0)$, write $(z=D_4)$. The internal Gram/Baikov
polynomial becomes
\begin{align}
G_{\rm cut}(z)=x_0^2x_2^2-\left(z-\frac \bc 2+x_2^2\right)^2\,,   
\end{align}
so its two branches are
\begin{align}
z_\pm=\frac \bc 2-x_2^2\pm x_0x_2\,.
\end{align}
Since $I^{110,-1}$ contains one $D_4$-numerator relative to
$I^{110,0}$, their leading singularities obey
\begin{align}
\frac{\operatorname{LS}_\pm(I^{110,-1})}
{\operatorname{LS}_\pm(I^{110,0})}=z_\pm\,.
\end{align}
Thus the branch-adapted combinations are
\begin{align}
N_+=I^{110,-1}-z_+ I^{110,0},\qquad
N_-=I^{110,-1}-z_- I^{110,0}\,.
\end{align}
They diagonalize the homogeneous $(2\times2)$ top-sector block of $A$ at
$\delta=0$. At non-zero $\delta$, both off-diagonal entries in this
basis are generally nonzero but proportional to $\delta$. Therefore the
leading-singularity change of basis by itself does not give an exactly
triangular finite-$\delta$ connection.

We may then extract some insight into the class of function that appear in the integration of our toy model by inspecting the entries of the two bottom (top sector) rows of the triangular connection $\mathbf{A}$. First we note that 
\begin{align}
    A_{j,2}=0, j\neq 2
\end{align}
So that $ \bar I^{010;0}(\ba)$ does not contribute to the other sectors. We still keep this vector since it will contribute to the reduction of generic $\bar I^{n_1n_2n_3;0}$ for $\delta>0$. Similarly, $A_{54}=0$. The remaining $A_{4j}$ and $A_{5j}$ are rational functions with simple poles. In particular, the top sector block diagonalizes into
\begin{align}
    \begin{pmatrix}
    -\frac{1}{2}\partial_a\log \tilde\xi_0&0\cr
    0&-\frac{1}{2}\partial_a\log\tilde\kappa_0
    \end{pmatrix}\,,
    \end{align}
    where
\begin{align}\label{eq:kappaxi}
    \tilde\kappa_0&=4\lambda_0 + 4 \ba (x_0 - x_2)^2 + 4 \bb x_2^2+ 4 \bc  x_2(x_0 - x_2) \nonumber\\
 \tilde\xi_0&=4\lambda_0  + 4 \ba (x_0 + x_2)^2+ 4 \bb x_2^2 - 4 \bc x_2 (x_0 + x_2)\,.
\end{align}
We thus conclude, in particular, that this contribution to $\bar I^{110}(\ba)$ is at most a logarithm of degree 2 rather than 4, as might naively be expected. The explicit expressions for the remaining rational $A_{ij}$ will not be relevant for the further computation but are contained in the Mathematica package [\href{https://github.com/IvoSachs/in-in_correlators-}{CosmologicalBaikovRemainingMastersEpsilonZero.nb}]. 

\subsection{Complete Toy model basis}
We now include the remaining terms on the r.h.s. of \eqref{eq:diftoy} in the reduction. For brevity we will be less explicit. The complete  Mathematica package for this reduction is contained in [CosmologicalBaikov111DaComplete.nb]. We found a complete set of basis integrals for the complete reduction of the form
\begin{align}
    \mathbf{M}=
\left(
I^{000,0},I^{010,0},I^{100,0},I^{001,0},I^{110,0},I^{110,-1},I^{101,0},I^{101,-1},I^{011,0},I^{011,-1},I^{111,0}
\right)^T .
\end{align}
The $I^{101}$ connection follows from $I^{011}$ by the permutation
\begin{align}
(a,b,c,x_2^2,x_3^2)\mapsto
(a+b-c,b,2b-c,x_0^2,x_3^2)\,,    
\end{align}
together with
\begin{align}
   D'_4=D_3-D_4+b-x_3^2\,. 
\end{align}
The final row of the connection reads
\begin{align}
    A_{111,\bullet}=
\frac{
 \tfrac12 e_{110}^{T}
 +(a-\tfrac c2)A_{110,\bullet}
 +(b-\tfrac c2)A_{101,\bullet}
 +\tfrac c2 A_{011,\bullet}
}{\widetilde\Lambda_0}
+\frac{b-x_3^2}{2\widetilde\Lambda_0}e_{111}^{T}\,,
\end{align}
where 
$e_{110}^{T}$ and $e_{111}^{T}$ are the projectors on the $I^{110}$ and $I^{111}$ masters respectively. 
Thus no additional master integral is needed for $\partial_a I^{111}$.
At $\delta=0$, the three two-master sectors are transformed using
\begin{align}
    P_{ij}=
\begin{pmatrix}
-z^{ij}_+&1\\
-z^{ij}_-&1
\end{pmatrix}\,,
\end{align}
with
\begin{align}
z^{110}_\pm&=\frac c2-x_2^2\pm x_0 x_2\,,\qquad z^{101}_\pm=\frac c2-x_3^2\pm x_0 x_3\,,\qquad z^{011}_\pm=\frac c2\pm x_2x_3\,.
\end{align}
The resulting complete $11\times11$ connection is lower triangular:
\begin{align}\label{eq:toycon}
    \begin{pmatrix}
    \cdots&\cdots&\cdots&\cdots&\cdots&\cdots&\cdots\cr\cdots&
    -\frac{1}{2}\partial_a\log\tilde\xi_0&0&\cdots&0&0&0\cr
   \cdots &0&-\frac{1}{2}\partial_a\log\tilde\kappa_0&\cdots&0&0&0\cr\cdots&0&0&\cdots
&0&0&0\cr\cdots&0&0&\cdots&-\frac{1}{2}\partial_a\log\tilde\gamma_0&0&0\cr\cdots&0&0&\cdots&0&-\frac{1}{2}\partial_a\log\tilde\beta_0&0&\cr
\cdots&\frac{-1+(\ba-\frac{\bc}{2})\partial_a\log\tilde\xi_0}{4x_0x_2\tilde\Lambda_0}
&
\frac{1+(\frac{\bc}{2}-\ba)\partial_a\log\tilde\kappa_0}{4x_0x_2\tilde\Lambda_0}
&\cdots
     &\frac{\bc \partial_a\log\tilde\gamma_0}{8\tilde\Lambda_0x_2x_3}& -\frac{\bc\partial_a\log\tilde\beta_0}{8\tilde\Lambda_04x_2x_3}& -\frac{1}{2}\partial_a\log \tilde\Lambda_0\end{pmatrix}\,,
\end{align}
where
\begin{align}
 \tilde\beta_0&=\bc^2 + 4 \bb x_2^2 + 4 \bc x_2 x_3 + 4 \ba (-\bb + x_3^2)\nonumber\\
 \tilde\gamma_0&=\bc^2 + 4 \bb x_2^2 - 4 \bc x_2 x_3 + 4 \ba (-\bb + x_3^2)\,.
\end{align}
We can easily recognize the structure of eqn. \eqref{eq:diftoy} in the connection matrix \eqref{eq:toycon}. The source terms for $\bar I^{111}$ in the last line \eqref{eq:toycon} are such that the poly-log degree of $\bar I^{111}$ is the same as for $\bar I^{110}$. Combining this with the discussion above \eqref{eq:kappaxi} we conclude that  $\bar I^{110}$ is a di-logarithm. Including the subtopologies of the flat space triangle in \eqref{eq:flat-connection} will not change this result since they are given by simple logarithms. In the next section, for completeness, we will nevertheless include the complete flat space triangle and derive the corresponding master basis and connection matrix. 

\subsection{Complete Triangle}\label{sec:ctr}
We finally include the sub-topologies of the flat space triangle in \eqref{eq:flat-connection}. As explained above they will not affect the poly-log degree of $I^{111}$ given that they integrate to simple log's they can at most contribute as a di-log to $I^{110}$ etc. They do, of course affect the concrete form, or symbol \cite{ Pimentel:2026kqc,Chowdhury:2026dwm,Henn:2026lfz} of the final result. 

If we include sub-topologies of the flat space triangle the first reduction equation acquires  addition terms, 
\begin{align}\label{eq:diftoyc}
   \tilde\Lambda_0\partial_{\ba} \bar I^{111}=& \frac{1}{2}(\bb-x_3^2
   )I^{111}+\frac{1}{2}I^{110}+(\ba-\frac{\bc}{2})\partial_{\ba}I^{110}+ (\bb-\frac{\bc}{2})\partial_{\ba} I^{101}+\frac{\bc}{2}\partial_{\ba} I^{011}\,,\nonumber\\
   &+\frac{1}{2}\int i_{111}\left(\frac{Q_2\cdot Q_3}{Q_2^2}(T_{101}-T_{011})+\frac{Q_3^2-Q_2\cdot Q_3}{(Q_2-Q_3)^2}(T_{101}-T_{110})\right)\,.
   \end{align}
Our Ansatz for the complete basis of master integrals is then given by 
\begin{align}
    \bm M_R+\bm M_{x}+\bm M_{y}+\bm M_{\bm T}\,,
\end{align}
where $\bm M_{\bm T} =\bm M\times  T_{111}$ is identical with that of the complete toy model multiplied by the top sector of the flat space triangle. On the other hand, acting with $\partial_{\ba}$ on the r.h.s. of \eqref{eq:diftoyc} produce finite sub--topologies of the bubble graphs with no $\Lambda$ in the denominator, as can be seen, for instance by differentiating elements of $\bm f$ in \eqref{eq:f-basis}, i.e. 
\begin{align}\label{eq:daL}
    \partial_{\ba} L_x=\frac{1}{D_{10}}-\frac{1}{D_{5}}\,,\quad \partial_{\ba} L_y=\frac{1}{D_{10}}\,,
\end{align}
where, $D_{10}=(Q_2-Q_3)^2$ was introduced at the beginning of this section. We refer to them as the rational sector $\bm M_R$ as they correspond to diagrams with only auxiliary propagators without square-roots. To describe them it is convenient to  extend the collection \eqref{tria4D1} by the tautological element $D_{10}$,
\begin{align}
&I^{m_1m_2m_3r_1r_2r_3r_4;j}_{n_1n_2n_3}= \int\limits_{-\infty}^\infty \frac{[dp_2]_\delta [dp_3]_\delta}{D_1^{m_1}D_2^{m_2}D_3^{m_3}D_4^{r_1}D_5^{r_2}D_6^{r_3}D_{10}^{r_4}\Lambda^j }\int  \frac{d^{4-2\epsilon}L}{D_7^{n_1} D_8^{n_2} D_9^{n_3}}\,.
\end{align}\label{tria4D1e}
An ordered basis for $\bm_R$ is then (with $I^{\bullet,\bullet}_{\bullet}:=I^{\bullet,\bullet;j=-1/2}_{\bullet}$)
\begin{align}
    \bm M_R= \{ &I^{110,-;-}_{-},I^{100,0100}_{-},I^{010,0001}_{-},I^{-,0101}_{-},I^{101,-}_{-},I^{001,0001}_{-},I^{011,-}_{-},\cr
&\qquad\qquad I^{001,0100}_{-},I^{111,-}_{-},I^{101,0100}_{-},I^{011,0001}_{-},I^{001,0101}_{-}\}\,,
\end{align}
where "-" stands for a row of zeroes. Finally, the finite bubble sub-topologies of of the flat space triangle $L_y$ and $L_x$ combine with a four-dimensional subset of $\bm M_R$ into 
\begin{align}
    \bm M_{x/y}=\{ I^{110,-}_{-}\times L_{x/y}\,, I^{101,-}_{-}\times L_{x/y}\,,I^{011,-}_{-}\times L_{x/y}\,,I^{111,-}_{-}\times L_{x/y}\}
\end{align}
Altogether, the ordered basis 
\begin{align}
 \bm M=   (\bm M_R, \bm M_x, \bm M_y,\bm M_{\bm T})^T
\end{align}
 is $12+4+4+11=31$-dimensional. In this filtration, the corresponding connection matrix 
 \begin{align}
     \partial_{\ba} \bm M= \bm A \bm M\,,\quad \bm A=\begin{pmatrix}
         \bm A_R&0&0&0\cr \bm C_{yR}&A_y&0&0\cr
         \bm C_{xR}&0&A_x&0\cr
         0& \bm C_{\bm T y}&\bm C_{\bm T }& \bm A_{\bm T}
     \end{pmatrix}\,,
 \end{align}
 takes a block triangular form. The explicit $4\times 12$-matrix follow from \eqref{eq:daL}. The block $\bm A_{\bm T}$ is identical with \eqref{eq:toycon} of the complete toy model and can brought into lower triangular form in the same way as described in section \ref{sec:diag}. Finally, the rational sector is already triangular. The explicit form of $\bm A$ is contained in the Mathematica notebook [\href{https://github.com/IvoSachs/in-in_correlators-}{CosmologicalBaikovFiniteNestedBasis.nb}]. We note, in closing, that the analysis presented so far does not determine the cosmological triangle completely. Nor does it contain the complete information about the alphabet \cite{Goncharov:2010jf}. To do so we would need to repeat the above to the remaining differential equations w.r.t. $\bb,\bc$ and $x_i$. The former can be obtained from the results above by a permutation, while the latter reveal further information about the boundary conditions. We will not pursue this further as the aim of this paper is not to provide an explicit solution for the triangle which has prviously been obtained in \cite{Pimentel:2026kqc}. 

\section{Tree-level Exchange}
As an illustration of how differential equations and IBP's behave in the presence of more complicated auxiliary propagators we now consider a tree-level exchange diagram of a conformally coupled scalar field. For a flat space scattering amplitude this diagram does not involve any momentum integration, of course. However, for an $in\!-\!in$ correlator there is still an auxiliary propagator left. We then have
\begin{align}
\label{dressforD2}
D_4 &= \begin{tikzpicture}[baseline]
\draw (-0.75, 0.5) -- (-0.5, 0);
\draw (-0.75, -0.5) -- (-0.5, 0);
\draw (0.75, 0.5) -- (0.5, 0);
\draw (0.75, -0.5) -- (0.5, 0);
\draw (-0.5,0) -- (0.5,0);
\draw[dashed] (0.5, 0) -- (0, -0.75);
\draw[dashed] (0, -0.75) -- (-0.5, 0);
\node at (0, -0.75) {$\Circle$};
\node at (-0.5,0) {$\bullet$};
\node at (0.5,0) {$\bullet$};
\node at (-1, 0) {$x_1$};
\node at (1, 0) {$x_2$};
\node at (0, 0.35) {$ \vec{y}$};
\end{tikzpicture}  \\
&=-\frac{1}{\pi}\int_{x_1}^{\infty} ds_1 \int_{x_2}^{\infty} ds_2\int\limits_{-\infty}^{\infty}dp\frac{p^2 }{(p^2 + s_1^2 ) (p^2 + s_2^2 )(p^2 +y^2)},\nonumber
\end{align}
where $y^2=(\vec{k}_1+\vec{k}_2)^2$, and $x_1= |\vec{k}_1|+|\vec{k}_2|$, $x_2= |\vec{k}_3|+|\vec{k}_4|$. However, it may be better to go "off-shell" where $x_i$ are independent variables. On the physical sheet, where $x_i,y>0$, the result of the integrals is given by a sum of di-logarithms \cite{Arkani-Hamed:2015bza}:
\begin{align}
\label{D2explicit}
    D_4 = \frac{1}{2 y} &\bigg( \text{Li}_2\big( \frac{x_1-y}{x_1+x_2}\big)+\text{Li}_2\big( \frac{x_2-y}{x_1+x_2}\big)+\log\big(\frac{x_1+y}{x_1+x_2}\big) \log\big(\frac{x_2+y}{x_1+x_2}\big)-\frac{\pi^2}{6}\bigg)\,.
    \end{align}
Below we will show how this result follows from the same differential equation as in the previous sections.\footnote{There is second diagram with dotted propagators that contributes as $\frac{\pi^2}{y}$ to this process \cite{Chowdhury:2025ohm}. We will not consider it here since it just affects the boundary condition.   } 
\subsection{Differential equations}
For brevity we will focus on the kinematical variable $y^2$. Derivatives w.r.t. $x_i^2$ can be treated in a similar manner. 
We have 
\begin{align}
    y^2\partial_{y^2} D_4&=-D_4 -\frac{1}{\pi}\int_{x_1}^{\infty} \!\!ds_1 \!\int_{x_2}^{\infty} \!\!ds_2\!\int\limits_{-\infty}^{\infty}dp\frac{p^2\;p^2 }{(p^2 + s_1^2 ) (p^2 + s_2^2 )(p^2 +y^2)^2},\nonumber\\
    =&\frac{1}{2}D_4 +\frac{1}{\pi}\!\int_{x_1}^{\infty}\!\!\! \!   ds_1\!  \int_{x_2}^{\infty}\! \!  \! ds_2\! \! \int\limits_{-\infty}^{\infty}\! dp\! \left(\frac{p^2 }{(p^2 + s_1^2 )^2 (p^2 + s_2^2 )}+ \frac{p^2 }{(p^2 + s_1^2 ) (p^2 + s_2^2 )^2}\right)\frac{\;p^2}{(p^2 +y^2)},\nonumber\\
    =&-\frac{3}{2}D_4 -\frac{1}{\pi}\!\int_{x_1}^{\infty} \! \! \! \! ds_1 \! \int_{x_2}^{\infty}\! \! \!  ds_2\! \! \! \int\limits_{-\infty}^{\infty}\! dp\left(\frac{s_1^2 }{(p^2 + s_1^2 )^2 (p^2 + s_2^2 )}+ \frac{s_2^2 }{(p^2 + s_1^2 ) (p^2 + s_2^2 )^2}\right)\frac{\;p^2}{(p^2 +y^2)},\nonumber\\
    &=-\frac{1}{2}D_4 -\frac{1}{2\pi}\left(\int_{x_2}^{\infty} ds_2\int\limits_{-\infty}^{\infty}dp\frac{x_1 }{(p^2 + x_1^2 ) (p^2 + s_2^2 )}\right.\nonumber\\&\qquad\qquad\qquad\qquad\left.+ \int_{x_1}^{\infty} ds_1\int\limits_{-\infty}^{\infty}dp\frac{x_2}{(p^2 + s_1^2 ) (p^2 + x_2^2 )}\right)\frac{\;p^2}{(p^2 +y^2)}\,,
\end{align}
where we have used IBP for $p$ in the second equality, and IBP for $s_1$ ($s_2$) in the 4th equality. To continue we use partial fractions to write the inhomogeneity in the last line as 
\begin{align}
   &\frac{x_1/2\pi}{y^2-x_1^2}\int_{x_2}^{\infty} ds_2\int\limits_{-\infty}^{\infty}dp\left(\frac{p^2 }{(p^2 + s_2^2 ) (p^2 + x_1^2 )}-\frac{p^2 }{(p^2 + s_2^2 ) (p^2 + y^2 )}\right)\nonumber\\
   &\qquad\qquad+\frac{x_2/2\pi}{y^2-x_2^2}\int_{x_1}^{\infty} ds_1\int\limits_{-\infty}^{\infty}dp\left(\frac{p^2 }{(p^2 + s_1^2 ) (p^2 + x_2^2 )}-\frac{p^2 }{(p^2 + s_1^2 ) (p^2 + y^2 )}\right)\nonumber\\
   &   \equiv \frac{x_1}{y^2-x_1^2} L_{x_1y}+ \frac{x_2}{y^2-x_2^2}L_{x_2y}\,.
\end{align}
Since $L_{x_1y}$ and $L_{x_2y}$ are easily integrated explicitly we will not bother to decompose them into further sub-topologies. We thus have two types of masters, 
\begin{align}\label{eq:lxy}
  L_{xy}= \quad&\begin{tikzpicture}[baseline]
\draw (-1,0) -- (1,0);
\node at (-1,0) {$\bullet$};
\node at (1,0) {$\bullet$};
\node at (0,.7) {$x$};
\draw[dashed] (1,0) arc(0:180:1cm and 0.5cm);
\end{tikzpicture}\quad=\frac{1}{2}\log(\frac{x+y}{x_1+x_2})\quad\text{and}\nonumber\\\nonumber\\
 D_4=\quad  &\begin{tikzpicture}[baseline]
\draw (1,0) -- (-1, 0);
\draw[dashed] (0,1) -- (1, 0);
\draw[dashed] (0,1) -- (-1, 0);
\node at (0.6,0.8) {$x_2$};
\node at (-0.6,0.8) {$x_1$};
\node at (-1,0) {$\bullet$};
\node at (1,0) {$\bullet$};
\node at (0,1) {$\Circle$};
\end{tikzpicture}\,,
\end{align}
which satisfy the differential equation 
\begin{align}
    y^2\partial_{y^2}
\begin{pmatrix}
\cdots\\
\cdots\\
L_{{x_2y}}\\
L_{{x_1y}}\\
D_4
\end{pmatrix}
=\begin{pmatrix}
\cdots&\cdots&0&0&0\\
\cdots&\cdots&0&0&0\\
\cdots&\cdots&0&0&0\\
\cdots&\cdots&0&0&0\\
0&0&\frac{x_1}{x_1^2-y^2}&\frac{x_2}{x_2^2-y^2}&-\frac{1}{2}
\end{pmatrix}\begin{pmatrix}
\cdots\\
\cdots\\
L_{{x_2y}}\\
L_{{x_1y}}\\
D_4
\end{pmatrix}\,.\end{align}
Then, using the explicit form of $L_{xy}$ in \eqref{eq:lxy} as well as  $y^2\partial_{y^2}=\frac{y}{2}\partial_y$, it is not hard to see that the integral of the last line, with $L_{xy}$ as a source term reproduces \eqref{dressforD2} when supplemented with the correct boundary condition. 

\section{Conclusions} 
In this paper we formulated differential equations for dressed cosmological correlators at future infinity of deSitter space. They are very close, in nature to their higher loop flat space cousins. For instance the cosmological tree-level exchange maps to a flat space triangle while the cosmological triangle gets mapped to a 3-loop diagram in flat space which is already a complicated system. However, since the auxiliary energy integrals do not involve, spatial external momenta the corresponding IBP's simplify considerably with the help of partial fractions throughout the reduction. The standard reduction packages available do not make full use of this simplicity, but can easily be adapted to do so. Here, we considered a few examples to illustrate the reduction mechanism. The connection matrix allows to understand some peculiar features of these correlators such as the di-log nature of the triangle \cite{Pimentel:2026kqc,Henn:2026lfz}. In particular, our toy model where we replace the flat space triangle by its leading singularity, is sufficient to infer the poly-log degree of the final result. For the latter the basis of master integrals is quite small consisting of just 11 elements rendering the analysis quite manageable. 
The factorization of the correlator into flat space and auxiliary (outer sector) is a distinguishing feature of the dressing rules. It has recently been used in \cite{Henn:2026lfz} to simplify the impressive explicit calculation of the cosmological triangle \cite{Pimentel:2026kqc}.  

The reduction of the flat space contribution to its leading singularity can also be used for more complicated diagrams such as the cosmological box diagram for instance. This already shows that the latter is in the elliptic class (see also \cite{Schwartz:26}). 
Having a cosmological version of the standard flat space packages would provide a helpful tool to push this analysis to more complicated integrals. Also, while here we have used a variant of the Laporta-IBP-master integrals formulation, alternative approaches such as using annihilators (e.g. \cite{Lairez:2022zkj}) might be more effective. 

There is no obvious obstruction to extend this formulation to higher loop such as the cosmological ice-cream \cite{Heckelbacher:2022hbq,Chowdhury:2023arc}, or massless- and  massive fields. For the latter case it would be interesting to analyze if observable cosmological such as local- and non-local signals \cite{Arkani-Hamed:2015bza} can be extracted from the connection matrix.  

\section*{Acknowledgments:}
I would like to thank Chandramouli Chowdhury, Christoph Nega and Lorenzo Tancredi for early collaboration and their invaluable help throughout this project. Without their help and insight on differential equations this paper could not have been completed. Furthermore, I would to like thank Johannes Henn and Qinglin Yang for sharing their  results with me prior to publication. ChatGTP-5.6 Sol was used for the generation of the Mathematica notebooks for the IBP reduction. This work was supported in parts by the Excellence
Cluster Origins of the DFG under Germany’s Excellence Strategy EXC-2094 390783311. 
\bibliographystyle{JHEP}
\bibliography{references.bib}
\end{document}